\documentclass[traditabstract]{aa} 
\usepackage{times,amssymb,amsmath} 
\usepackage{epsfig} 
\usepackage{lscape}

\begin{document} 

\title{Stellar properties of $z\sim1$ Lyman-break galaxies \\ from ACS slitless grism spectra}

\author{K.K. Nilsson\inst{1}
        \and O. M{\"o}ller-Nilsson\inst{2}
        \and P. Rosati\inst{1}
        \and M. Lombardi\inst{1}
        \and M. K{\"u}mmel\inst{1}
        \and H. Kuntschner\inst{1}
        \and J.R. Walsh\inst{1}
        \and R.A.E. Fosbury\inst{1}
}

\institute{
   ST-ECF, Karl-Schwarzschild-Stra\ss e 2, 85748, Garching bei M\"unchen, Germany\\	
\and
   Max-Planck-Institut f{\"u}r Astronomie, K{\"o}nigstuhl 17,
   69117 Heidelberg, Germany\\
}
\offprints{knilsson@eso.org}
\date{Received date / Accepted date}
\titlerunning{ACS slitless grism observations of $z\sim1$ LBGs}

\abstract
{Lyman-break galaxies are now regularly found in the high redshift Universe by searching for the break in the galaxy spectrum caused by the Lyman-limit redshifted into the optical or even near-IR. At lower redshift, this break is covered by the \emph{GALEX} UV channels and small samples of $z\sim1$ LBGs have been presented in the literature. Here we give results from fitting the spectral energy distributions of a small sub-set of low redshift LBGs and demonstrate the advantage of including photometric points derived from HST ACS slitless grism observations. The results show these galaxies to have very young, star forming populations, while still being massive and dusty. LBGs at low and high redshift show remarkable similarities in their properties, indicating that the LBG selection method picks similar galaxies throughout the Universe.}


\maketitle

\section{Introduction}
To understand galaxy evolution we need to understand the stellar properties of galaxies at high redshift. These galaxies include samples of mainly Lyman-break galaxies (LBGs; e.g. Steidel et al. 1996, 1999, Pettini et al. 2001, Shapley et al. 2003, Bunker et al. 2004, Ouchi et al. 2004, Burgarella et al. 2007), but also other types of galaxies such as Ly$\alpha$ emitters (LAEs; e.g. M{\o}ller \& Warren 1993, Fynbo et al. 2002, Venemans et al. 2007, Nilsson et al. 2009a) or very red and massive galaxies (DRGs, EROs; Franx et al. 2003, Daddi et al. 2004). One common method of studying the stellar populations of these galaxies is by fitting their spectral energy distributions (SEDs) with spectral templates, thereby determining luminosity-weighted properties such as stellar ages, dust content, metallicities, and stellar masses. 

For Lyman-break galaxies, the last decade has seen a large number of publications determining the stellar properties using SED fitting. Two of the first publications, Shapley et al.~(2001) and Papovich et al.~(2001) presented results for LBGs at $z=2-3$. They found that LBGs at these redshifts tended to be young, with ages of a few hundred Myrs respectively, have masses around $10^{10}$~M$_{\odot}$ and significant extinction of the order of A$_V = 0.7 - 1.2$~mag. The star formation rates (SFRs) in their samples were a few times 10~M$_\odot$~yr$^{-1}$. SED fitting results at higher redshifts seem to indicate some evolution in these properties. Verma et al.~(2007) and Yabe et al.~(2009) fit LBGs at $z \sim 5.5$ and find that they are typically younger (a few times 10~Myrs), less massive (a few times $10^9$~M$_{\odot}$), and less dusty ($A_V = 0.3 - 0.7$~mag.). They also find that the higher redshift LBGs have higher SFRs, of the order $50 - 100$~M$_\odot$~yr$^{-1}$, indicating that these are small galaxies in the process of building up their stellar mass. The evolution seen from smaller, younger galaxies at $z\sim5$ to larger and less vigorously star-forming galaxies at $z \sim 2.5$ does not seem to continue to even lower redshift. Overzier et al.~(2009) presented a sample of local ``LBG analogues'' based on UV bright galaxies in SDSS. These galaxies have masses of $6 \times 10^9$~M$_{\odot}$ and SFRs of $\sim 10$~M$_\odot$~yr$^{-1}$, i.e. they are of similar mass and have similar SFRs to LBGs at $z \sim 2.5$.

Whereas SED fitting remains a popular method to determine stellar properties of  LBGs, potential caveats with this method are the lack of a good wavelength spread in, or numbers of, spectral points and low signal-to-noise in the observations. In this paper we aim to avoid both these potential sources of uncertainty by observing a sample of pre-selected, lower redshift LBGs, at $z\sim 1$ (Burgarella et al.~2007). We also focus especially on LBGs with associated HST/ACS G800L grism spectra, since the spectra will, at these redshifts, completely encompass the Balmer and 4000~{\AA} breaks, thus allowing a very good sampling of these breaks. These breaks, in turn, are very effective indicators of the age of the stellar population and dust content (Hamilton 1985, Balogh et al. 1999, Kriek et al. 2006). In Kriek et al.~(2006) it is shown that not only the redshift, but also the stellar properties are better constrained when using spectroscopic observations covering the Balmer and 4000~{\AA} break of the galaxy, rather than just broad-band photometric points. Kauffmann et al.~(2003) show that stellar properties can be determined solely on the basis of the 4000~{\AA} break strength and the Balmer absorption-line index H$\delta_A$. Slitless grism observations with the HST G800L grism are particularly well suited to study these breaks in high redshift galaxies; with low wavelength resolution, but high sensitivity and low background. Here, we present the results of fitting 15 LBGs with grism spectra, and show that the stellar properties are very well constrained by these observations. 

In Sect.~\ref{sec:data} we describe the data used in the SED fitting, and in Sect.~\ref{sec:method} the fitting method is described. The results are given in Sect.~\ref{sec:results} and we end with a discussion in Sect.~\ref{sec:disc}. Throughout this paper, we assume a cosmology with $H_0=72$
km s$^{-1}$ Mpc$^{-1}$, $\Omega _{\rm m}=0.3$ and
$\Omega _\Lambda=0.7$. Magnitudes are given in the AB system. 

\section{Data}\label{sec:data}
\subsection{LBG sample}
The sample of LBGs used here is drawn from Burgarella et al.~(2007), who presented the selection of LBGs using GALEX observations in the redshift range of $z = 0.9 - 1.3$. Their sample includes 420 LBGs in the Chandra Deep Field South (CDFS). A sub-set of 32 of these lie in fields covered by public ACS slitless grism observations (K{\"u}mmel et al., in prep.). The ACS observations were made with grism G800L, with an effective wavelength range of $\sim 5\,500-10\,500$~{\AA} and a resolving power of 66 at 6365~{\AA} for point-sources. All public ACS slitless grism observations have been reduced by the PHLAG pipeline in an effort to provide science-ready grism spectra of almost $32\,000$ unique objects (K{\"u}mmel et al., in prep., but see also Pirzkal et al.~2004, Xu et al.~2007, Straughn et al.~2009 for similar observations). For this project, we specially extracted 1-D spectra from flux calibrated 2-D spectra for the 32 objects covered by grism observations, using a dedicated extraction script\footnote{See \\ \texttt{www.stecf.org/archive/hla/indiv\_extraction.php}}. The extraction provides error and contamination estimates for each wavelength pixel. Of the 32 spectra, four were not possible to retrieve due to their faintness or proximity to the edge of the CCD. Of the remaining 28, 15 were chosen for follow-up spectral fitting for having either a precisely known redshift from ground-based spectroscopy (13) or for having a very well-defined spectral break in the grism spectrum, allowing a redshift to be determined (2). The sample is defined in Table~\ref{tab:objects}. Ground-based spectroscopic redshifts were collected from all available public surveys in GOODS-S. Object LBG\_61554 had conflicting redshifts from VVDS ($z = 0.378$) and the K20 survey ($z = 1.087$). In this case we retain the K20 redshift as we see a break in the grism spectrum which cannot be explained with the lower redshift.
\begin{table*}[t]
\begin{center}
\caption{LBGs used in the fitting. }
\begin{tabular}{@{}lccccccclc}
\hline
\hline
LBG\_ID$_{UV}$ & RA & Dec & $z_{gb}$ & M$_V$  & FWHM & \#$_{spec}$ &  \#$_{points}$ & Grism\_ID \\
   & (deg)  & (deg) & & (mag)   & (arcsec) & &\\
\hline
46190  & 53.18483 & -27.92041 & ---  & $23.34$ & 10.0  & 4 & 39 & HAG\_J033244.36-275513.5\_J9FA43BIQ  \\
47362  & 53.17793 & -27.90950 &  $1.104^1$ & $22.46$  & 7.1  & 4 & 68 & HAG\_J033242.70-275434.2\_J9FA43BIQ  \\
49138 & 53.13903 & -27.89262 &  $0.679^2$ & $23.46$ & 6.1  & 1 & 18 & HAG\_J033233.37-275333.4\_J9FA43BIQ  \\
53929 & 53.03306 & -27.84798 &  $1.043^1$ & $23.79$ & 16.6  & 1 & 9 & HAG\_J033207.93-275052.7\_UDFNICP2   \\
58236 & 53.07825 & -27.80578 &  $0.999^3$ & $24.53$ & 12.7  & 2 & 9 & HAG\_J033218.78-274820.8\_J94SA3CIQ   \\
59257 & 53.17618 &	 -27.79613 &  $0.996^1$ & $22.39$ & 22.8 & 9 & 34 & HAG\_J033242.28-274746.1\_J8G6I3PEQ  \\
61554 & 53.14792 & -27.77405 & $1.087^4$ & $22.25$ & 34.3 & 10 & 16 & HAG\_J033235.50-274626.6\_J8G6I3PEQ  \\
61983 & 53.00104 & -27.77013 & $1.032^1$ & $23.70$ & --- & 2 & 14 & HAG\_J033200.25-274612.5\_UDFNICP2   \\
62478 & 53.01969 & -27.76527 & $1.079^5$ & $23.36$ & 16.4 & 1 & 5 & HAG\_J033204.72-274555.0\_UDFNICP2   \\
62522 & 53.06233 &	 -27.76517 & $1.226^2$ & $23.73$ & 12.9 & 1 & 15 & HAG\_J033214.96-274554.6\_J94SPDCSQ  \\
63294 & 53.12040 &	 -27.75701 & $0.953^5$ & $23.50$ & 14.9 & 5 & 39 & HAG\_J033228.90-274525.2\_J94SPDCSQ  \\
63984 & 53.10560 &	 -27.75079 & $0.973^1$ & $23.18$ & 6.3 & 4 & 56 & HAG\_J033225.34-274502.9\_J94SPDCSQ  \\
66235 & 53.04549 & -27.72864 & $0.998^3$ & $22.69$ & 29.7 & 3 &  6 & HAG\_J033210.92-274343.1\_J9FAA3IBQ  \\
69377 & 53.07893 & -27.69774 & --- & $22.74$ & 12.2 & 3 & 24 & HAG\_J033218.94-274151.9\_J9FAA3IBQ  \\
69919 & 53.04692 & -27.69087 & $1.059^5$ & $23.31$ & 7.0 & 4 & 64 & HAG\_J033211.26-274127.1\_J9FAA3IBQ \\
\hline
\label{tab:objects}
\end{tabular}
\end{center}
\begin{list}{}{}{}{}{}{}
\item[$^{\mathrm{}}$] IDs are from Burgarella et al.~(2007); RA and Dec (in J2000) from the grism position; M$_V$ is the isophotal magnitude from the GOODS-S catalogue v.2.0; FWHM are measured in the $z$ band from the GOODS-S catalogue. The last columns give number of overlapping grism spectra on each source (with different orientation angles), the number of extracted SED points for each object, and an example ID for the grism spectra. Notes on redshifts give the source of the ground-based redshift: 1) VVDS (Le F{\`e}vre et al.~2005), 2) IMAGES (Ravikumar et al.~2007), 3) FORS (Vanzella et al.~2008), 4) K20 (Mignoli et al.~2005) and 5) VIMOS (Balestra et al.~2010). 
\end{list}
\end{table*}

No particularly strong selection bias is expected in this sample, compared to higher redshift samples, since the original selection by Burgarella et al.~(2007) is expected to select the same type of galaxies as at higher redshift, and since the overlap with the grism spectra is a purely random sub-selection. The only bias may occur in the final selection from 28 objects with grism spectra, to the 15 used in the final analysis. Here, objects without ground-based spectra, typically fainter than the ones with ground-based spectra, are rejected. However, this bias may be considered mild, or insignificant, since the range of UV luminosities of the final sample of LBGs covers roughly the same absolute magnitude range, or even go a bit deeper, than the typical range at high redshift.

\subsection{Broad-band photometry}
To fit the SEDs of these galaxies, we need to collect photometric measurements from broad-band observations of the CDFS, and from the grism spectra. The photometry in the CDFS was done with aperture photometry on the images available in the GOODS-S HST (in filters F435W, F606W, F814W and F850LP, Giavalisco et al.~2004), the ESO ISAAC releases (Retzlaff et al.~2010), the ESO VIMOS $U$-band (Nonino et al.~2009), and the Spitzer IRAC $3.6$~$\mu$m band images. The HST images were convolved with a Gaussian function of FWHM $0.8''$ to match the ground-based data. The apertures in all images had a $2''$ diameter. One source (LBG\_61983) was not covered by the GOODS-S imaging, and the photometry for this source (in BV$iz$) was found in the MUSYC catalogues (Gawiser et al.~2006). Photometry in longer wavelength filters than the $3.6$~$\mu$m were not attempted as these correspond to $>1.5$~$\mu$m in the restframe where the spectral fitting models of our code are not very reliable. 

In order to test the accuracy of the photometry, specifically the agreement between photometry from different instruments, two tests were performed. Firstly, the broad-band photometry was run through the \emph{Hyperz} photometric redshift code (Bolzonella et al.~2000). Only the 14 galaxies with GOODS-S detections were tested, of which eight were assigned the correct redshift within the $1\sigma$ significance interval directly, for five the correct redshift was given as a secondary solution and only one (LBG\_53929) was assigned an incorrect redshift ($z\sim 2.3$). Hence, the photometry does not seem to suffer from any large sources of systematic error. Secondly, five unblended stars were found in the images and the procedure to extract photometry was repeated. The photometry from these stars were fitted with the stellar library of Gunn \& Stryker (1983). With the minimum error of 0.07~magnitudes used in the SED fitting here, the reduced $\chi^2$ of all five stars were for the best fits ranging between $1.79 - 3.10$. This test proves that there are no negative effects of potential aperture corrections, and that the minimum error bars chosen are optimally set.
 
\subsection{Grism photometry}
\begin{figure*}[!t]
\begin{center}
\epsfig{file=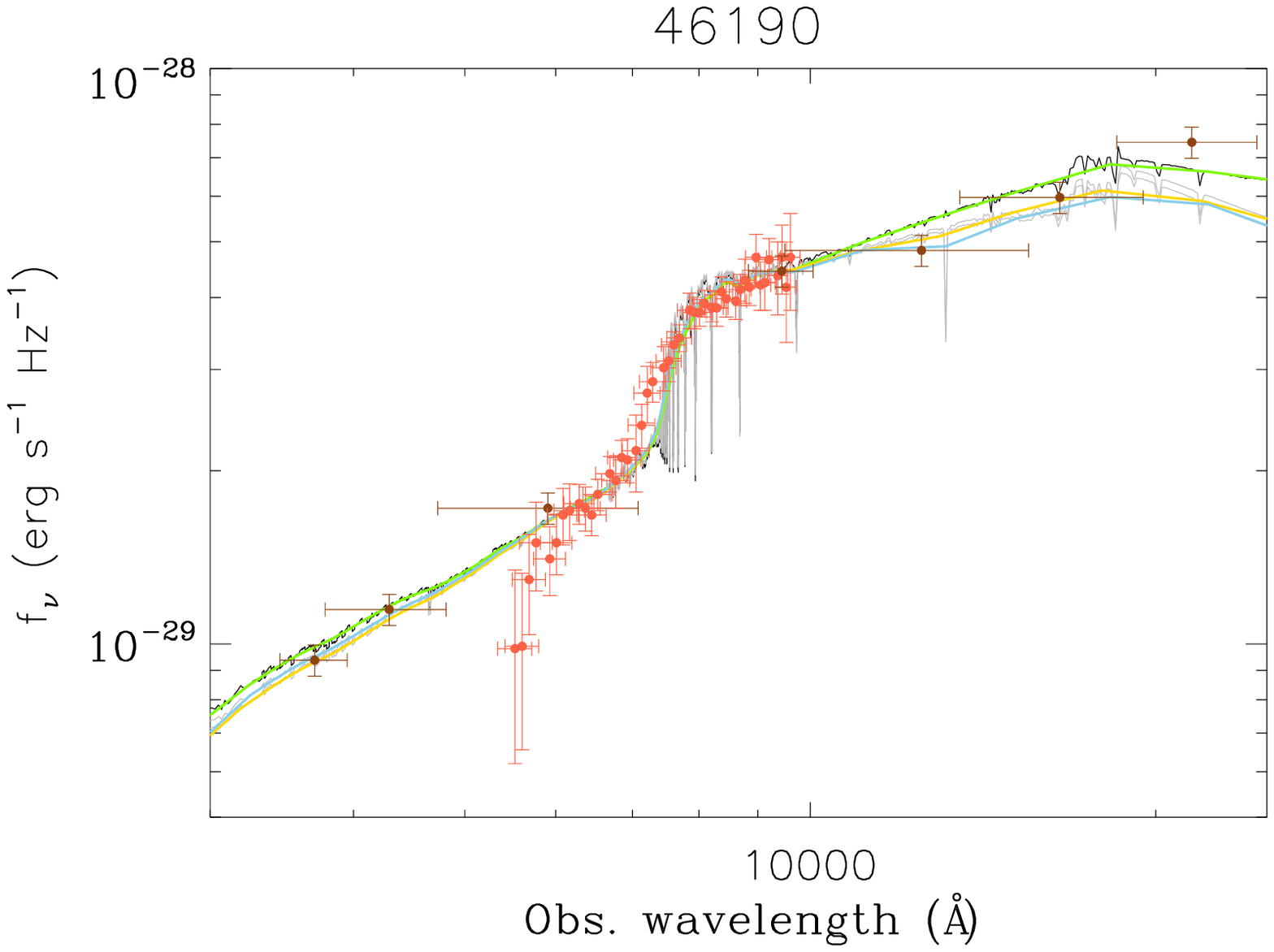,width=9cm}\epsfig{file=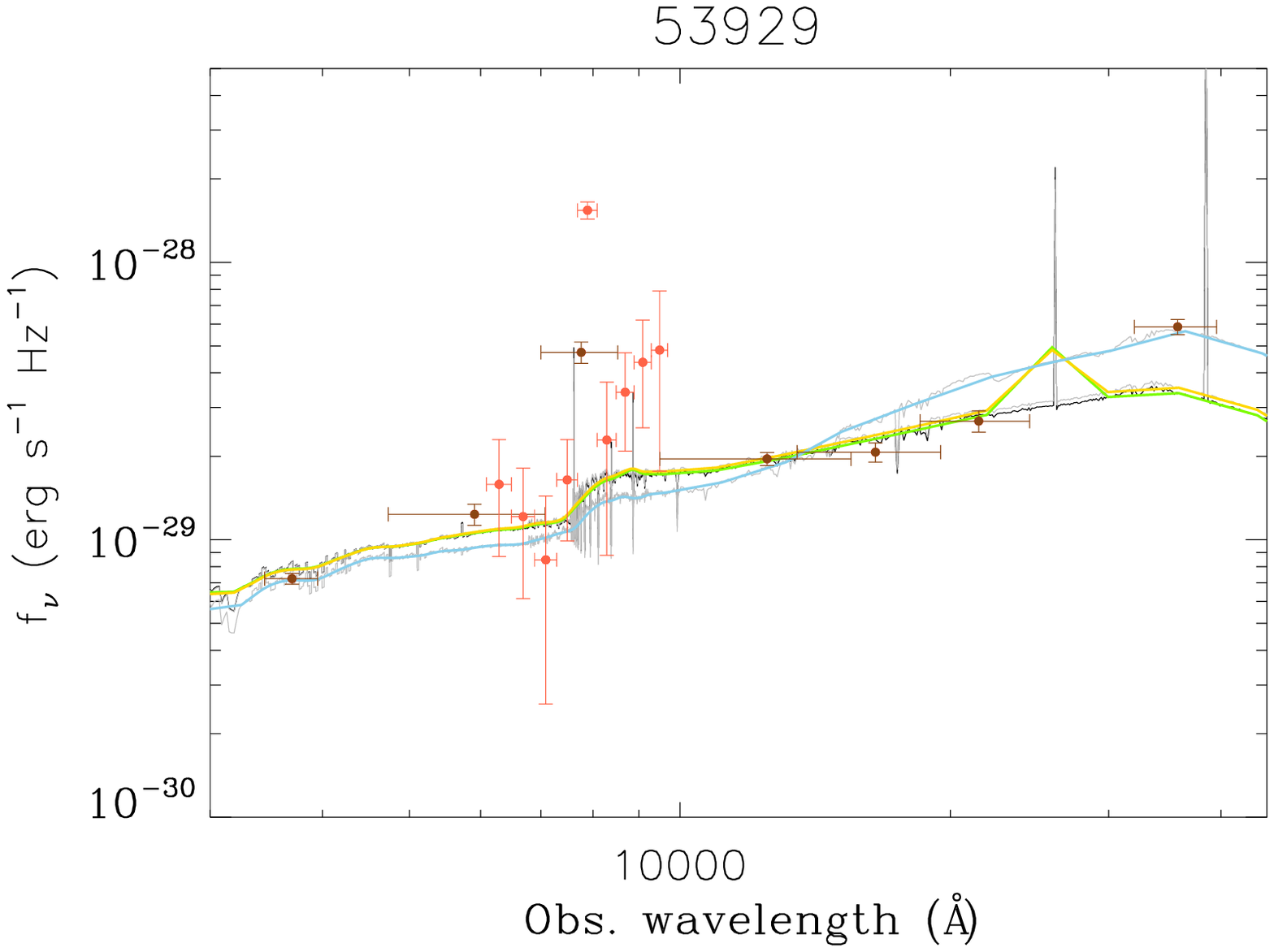,width=9cm}
\caption{SEDs and best fit spectra of two galaxies, one with a good $\chi^2_r$ and one with a worse fit. Points are photometric points, from broad-band photometry (brown) and from the grism spectra (red). Error bars in the wavelength direction show the FWHM of the filter with which the point was observed. The grey spectrum is the best fit spectrum from the models, with green, yellow and blue lines showing the spectra convolved with the resolution profile for each galaxy for the 1 SSP, 2 SSP and constant SFR scenarios.  LBG\_46190 (\emph{left}) has a $\chi^2_r \sim 1$ whereas LBG\_53929 has a $\chi^2_r \sim 25$ (\emph{right}). The brightest of the grism data points in the LBG\_53929 SED is due to a strong [OII] emission line. }
\label{fig:seds}
\end{center}
\end{figure*}

To constrain the Balmer break in these galaxies, photometry was extracted from the ACS slitless grism spectra in the following way. Spectral flux points in slitless grism spectra are correlated on small scales, and the correlation length along the wavelength direction depends on the size of the object. This is also manifested in a change in resolution with the size of the object. For a point-source object, the resolution of the grism is 66 at 6365~{\AA}, and the resolution scales linearly with object size according to:
\begin{equation}
R = 66 \times \frac{0.11}{PSW}
\end{equation}
where 0.11 is the intrinsic pseudo-slitwidth in arcseconds for a point-source, and $PSW$ is the pseudo-slitwidth in the dispersion direction of the actual object (as given by the \texttt{APERWID} keyword in the released spectra). The $PSW$ is in turn related to the morphological parameters of the object (i.e. major and minor axis, and position angle compared to spectral extraction direction). From the resolution we can calculate the FWHM and $\sigma$ of the dispersion. Extracting photometric points at $3\sigma$ intervals apart should then ensure virtually uncorrelated measurements of the SED of the object. Finally, for objects with multiple spectra at different HST/ACS G800L roll-angles, photometric points were extracted with $3\sigma$ intervals, but with slight offsets in wavelength, so that the extracted points sampled the entire spectrum more evenly. The extracted points were normalised to the broad-band photometry with the HST $i$ or $z$ magnitude, in order to eliminate effects caused by different aperture sizes, and then input into the SED fitting code. The magnitude used for the normalisation was excluded from the fit. Two examples of the SEDs are shown in Fig.~\ref{fig:seds}.

\section{SED fitting method}\label{sec:method}
For the fitting, the spectral models of GALAXEV were used (Bruzual \& Charlot 2003). The actual fits were made with the Monte-Carlo-Markov-Chain (MCMC) code NisseFit (Nilsson et al.~2007, Nilsson et al.~2010). We have described this code in some detail in previous publications, and will here only give a brief summary. The code uses a Metropolis Hastings algorithm to produce a Markov-Chain in a multi-dimensional parameter space. At each iteration point, the ``quality of fit'' parameter (in our case $\chi^2$)  is calculated by running the GALAXEV code with the given set of parameters and a Salpeter IMF. For single stellar population (SSP) models we add both nebular continuum and nebular emission lines based on age, metallicity and UV flux. These were calculated independently on a grid of the three parameters  using the SB99 (Leitherer et al.~1999)  and MAPPINGS (Kewley et al., in prep.) codes. The output spectrum is also corrected for the effect of dust according to Calzetti et al.~(2000) and redshifted. The model magnitudes in the different wavebands are extracted from the final spectrum and compared with the observed values to produce the ``quality of fit'' parameter, where we use a flat prior. 

For the fits here, from restframe $\sim 0.18 - 1.8$~$\mu$m, three runs were made with constant star formation rate (SFR), one SSP or two SSPs. The reason to fit SSPs is that we are only able to add nebular lines to these models, and two SSPs allow for a more realistic scenario to be tested. Single SSPs are then also run in order to compare the results, as many other studies have been made with single SSPs. A test of the validity of the SSP star formation history is then to compare the results to those of the runs with constant SFR. In all cases the age, metallicity, dust $A_V$ and stellar mass are free parameters. For constant SFRs, the SFR is also fitted, while for two SSPs we fit two ages and the mass fraction between the two populations. The minimum error on the broad-band measurements was set to 0.07 magnitudes to account for absolute calibration uncertainties. Redshifts were assumed to be those determined from the ground-based spectra. For the two without such values, the redshift was set to $z = 1.0$. Fits made with redshifts varied in increments of $z = 0.02$ between $z = 0.9 - 1.1$ all revealed similar results. Objects LBG\_49138 and LBG\_53929 have very strong emission-lines in their spectra~(see e.g.~Fig.~\ref{fig:seds}) that are fitted poorly by the SED code. In the first attempt to fit 2 SSPs, the respective reduced $\chi^2_r$ were 87 and 24. Manually removing 2/1 points in the grism photometry improved the fits to $\chi^2_r \sim 8$ for both galaxies. We report these results in the next section. 

\section{Results}\label{sec:results}
\subsection{SED fitting results}
\begin{figure*}[!t]
\begin{center}
\epsfig{file=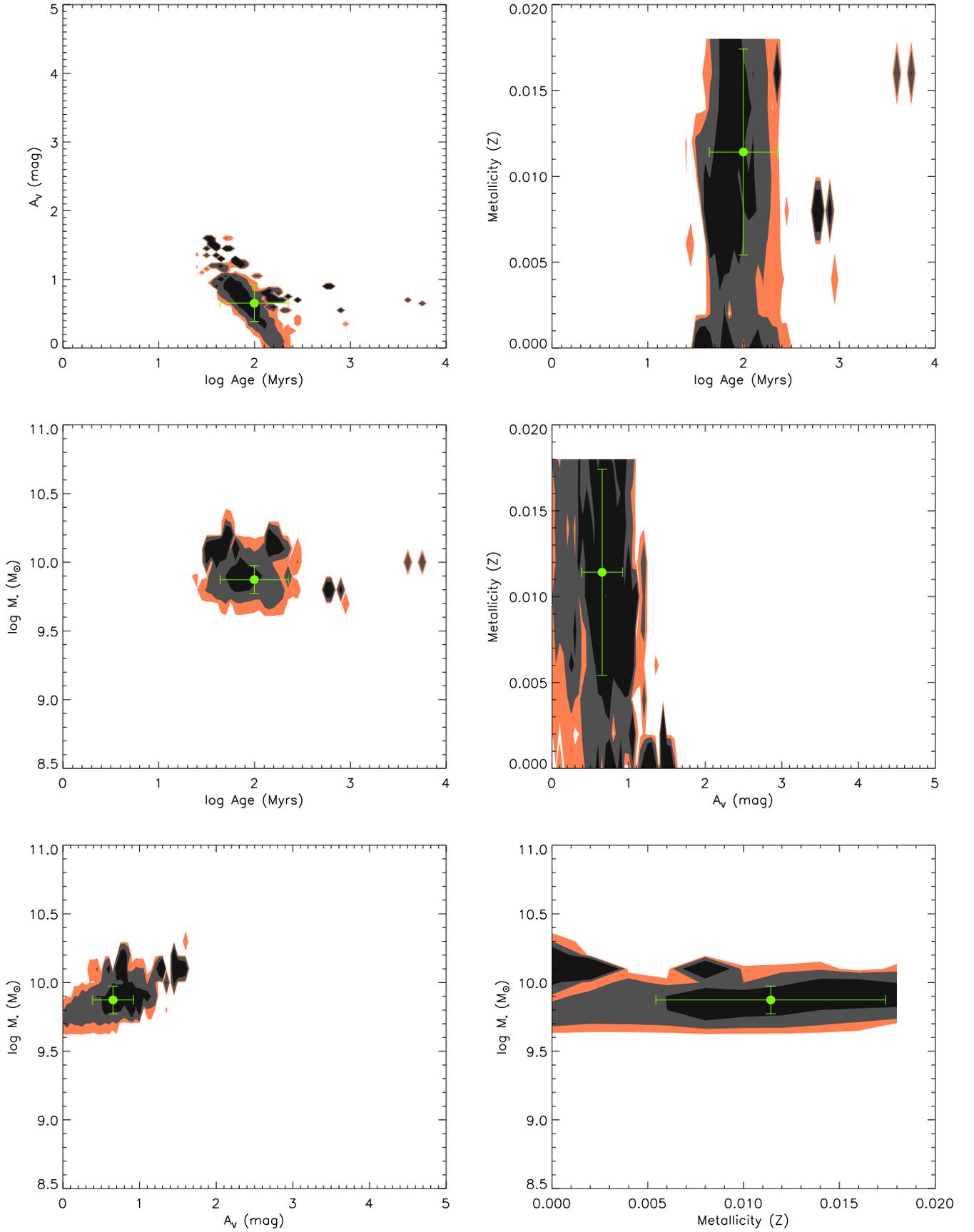,width=18cm}
\caption{Typical dependencies between parameters. From the 1 SSP fits to LBG\_46190. Contours show $1\sigma$,~$2\sigma$~and~$3\sigma$ confidence levels. The green point with error bars are the best fit 1D values, as given in Table~\ref{tab:results}. The metallicity is completely unconstrained, whereas the parameters A$_V$, stellar mass and age are relatively well constrained. There is a slight degeneracy between age and dust, with older ages requiring less dust.  }
\label{fig:params}
\end{center}
\end{figure*}

NisseFit was run $10\,000$ times for each galaxy, and the first $2\,000$ runs were excluded to reach the fits where the $\chi^2$ had converged to values near the minimum. The resulting runs then reflect the actual best fit distribution of the parameter probability space fitted. To study these probabilities, we collapse the runs into all possible 2-D combinations, and plot the significance contours. One example is shown in Fig.~\ref{fig:params}. It is clear that metallicity (in all fits) and SFR (in the constant SFR scenarios, not shown in Fig.~\ref{fig:params}) cannot be determined accurately, and we ignore these results hereafter. The parameters that are well determined are stellar mass, dust A$_V$ and age, although the age fits in the 2 SSP fits suffer from a degeneracy at the older age when the young population age is very young. In other words, when the young population age is very young, the light of these stars will outshine any older population, unless the mass fraction in the young population is very small (see Nilsson et al.~2010, for a longer discussion on this). The ages and mass fractions for all the 2 SSP fits (the only runs with more than one age fitted) indicate that we are not able to identify an older population in any of the objects, and only upper limits on the mass fraction in a potential old population may be inferred. 

To determine the best fit parameters in 1-D, a method similar to that of Nilsson et al.~(2009b) was used, where each parameter is decided individually and independently from the other parameters by finding the median of the distribution of runs and integrating to 68.3\% of each wing to find the scatter. The best fit results for the runs with two SSPs are found in Table~\ref{tab:results}. Here we only give the results of the runs with two SSPs, as the results for the constant star formation rate and 1 SSP runs agree very well with those for two SSPs.
\begin{table}[t]
\begin{center}
\caption{Results from the SED fits for the LBGs. }
\begin{tabular}{@{}lcccccc}
\hline
\hline
LBG\_id & Age & $A_V$  & log M$_{\star}$ & min $\chi^2_r$ & SFR \\
   & (Myrs)  & (mag) & (M$_{\odot}$) & & (M$_\odot$~yr$^{-1}$)\\
\hline
46190  & $68^{+19}_{-17}$  & $0.75^{+0.22}_{-0.23}$ & $10.06^{+0.20}_{-0.14}$ &  1.09 & $14\pm0.7$   \\
47362  & $21^{+13}_{-11}$  & $0.50^{+0.32}_{-0.38}$ &  $9.78^{+0.25}_{-0.24}$   &  2.07 & $42\pm2.0$ \\
49138 &  $5^{+3}_{-3}$        & $0.45^{+0.49}_{-0.31}$  &  $9.32^{+0.31}_{-0.24}$    & 8.24 & $5\pm0.2$ \\
53929 &  $33^{+2}_{-2}$      & $0.77^{+0.26}_{-0.30}$  & $9.67^{+0.34}_{-0.31}$    &  8.89 & $13\pm0.6$ \\
58236 &  $14^{+1}_{-1}$       & $1.15^{+0.07}_{-0.07}$ &  $9.10^{+0.22}_{-0.13}$   &  3.17 & $11\pm0.5$ \\
59257 &  $63^{+24}_{-20}$  & $1.10^{+0.23}_{-0.30}$ &  $10.56^{+0.25}_{-0.16}$ &  0.71& $39\pm1.8$ \\
61554 &  $24^{+18}_{-14}$ & $1.31^{+0.29}_{-0.31}$ &  $10.24^{+0.26}_{-0.20}$ &   2.69 & $64\pm3.0$ \\
61983 &  $70^{+18}_{-16}$ & $1.27^{+0.18}_{-0.24}$ &  $10.40^{+0.26}_{-0.13}$  &  0.42 & $23\pm1.1$ \\
62478 & $39^{+28}_{-24}$ &  $1.01^{+0.36}_{-0.28}$ &  $10.10^{+0.28}_{-0.24}$  &  2.06 & $25\pm1.2$ \\
62522 & $52^{+30}_{-30}$ &  $0.77^{+0.73}_{-0.48}$ &  $10.01^{+0.38}_{-0.30}$  &  9.21& $34\pm1.6$ \\
63294 & $62^{+23}_{-19}$ &  $0.45^{+0.18}_{-0.25}$ &  $9.73^{+0.24}_{-0.15}$ &    2.39 & $9\pm0.4$ \\
63984 & $38^{+16}_{-17}$ &  $0.24^{+0.41}_{-0.17}$ &  $9.48^{+0.26}_{-0.16}$ &    1.25 & $12\pm0.6$ \\
66235 & $48^{+20}_{-17}$ &  $2.06^{+0.25}_{-0.33}$ &  $10.87^{+0.29}_{-0.21}$ &  6.61& $64\pm6.2$ \\
69377 & $55^{+17}_{-14}$ &  $1.07^{+0.17}_{-0.19}$ &  $10.35^{+0.20}_{-0.11}$ &  3.58 & $29\pm1.4$ \\
69919 & $32^{+20}_{-18}$ &  $1.31^{+0.38}_{-0.44}$ &  $10.25^{+0.36}_{-0.23}$ &  1.16 & $42\pm4.1$ \\
\hline
\label{tab:results}
\end{tabular}
\end{center}
\begin{list}{}{}{}{}{}{}
\item[$^{\mathrm{}}$] Stellar properties from fitting 2 SSPs to the data. Columns give the age of the younger SSP, dust A$_V$, mass in stars, the minimum reduced $\chi^2$ and the dust corrected SFRs as calculated from the UV flux of all the fits. We here only give the age of the younger SSP, as any older stellar population is totally obscured by this younger population. LBGs 49138 and 53929 have been fitted after removing obvious strong emission lines. The bad $\chi^2$ of 62522 suffers from odd features in the NIR ISAAC photometry. Both object 53929 and 66235 lie on the edge of the HST imaging field and may thus also have larger systematic uncertainties on the photometry. Please note that the error bars in column six are the pure measurement errors on the U band photometry, and do not incorporate uncertainties in the extrapolation to 1500~{\AA}, or in the SFR conversion factor.
\end{list}
\end{table}
We see that most objects are fitted well, with reduced $\chi^2 \lesssim 3$. Objects with worse $\chi^2$ can be explained by features in the SEDs, from emission lines or too small error bars on the photometry due to systematic errors.

\subsection{Star formation rates}
From the U band photometry, probing restframe $\sim 1600-2150$~{\AA} at $z = 1$, we can calculate the UV-derived star formation rates of the galaxies. Using the conversion of Kennicutt~(1998), and extrapolating the flux in the U band to restframe $1500$~{\AA} assuming a flat continuum in $f_{\nu}$, the derived SFRs range from $2 - 21$~M$_\odot$~yr$^{-1}$, with a median value of $6.0$~M$_\odot$~yr$^{-1}$. These are the SFRs uncorrected for dust, which affects the UV light strongly. However, as the exact dust attenuation has been determined with the SED fitting, these can be corrected for the dust absorption using the same Calzetti et al.~(2000) law as in the actual SED fitting. After correction of the dust absorption, the dust corrected SFRs range from $5-64$~M$_\odot$~yr$^{-1}$, with a median SFR of $25.4$~M$_\odot$~yr$^{-1}$.

\section{Discussion}\label{sec:disc}
\subsection{Stellar properties of $z\sim1$ LBGs}
The results presented here come from fitting the SEDs of high redshift galaxies with theoretical templates. This fitting is hampered by a number of systematic uncertainties. Firstly, on the observational side, the inclusion of the grism spectrum in the photometric points helps constraining the fits very well, although for object LBG\_49138 and LBG\_53929, the strong [OIII] and [OII] emission lines had to be removed manually since they were difficult to fit accurately. One may further ask if the light from these galaxies may be contaminated by e.g. AGN. We searched the \emph{Chandra} X-ray 2~Ms catalogue (Luo et al.~2008) for counterparts of the galaxies and found only one, corresponding to LBG\_66235. The X-ray flux of this galaxy is $1.8\times 10^{42}$~erg~s$^{-1}$ in the $0.5-2$~keV band, corresponding to a SFR of 355~M$_\odot$~yr$^{-1}$ (Ranalli et al.~2003). This SFR is much higher than that derived from the UV (64~M$_\odot$~yr$^{-1}$) indicating that this galaxy may be a Seyfert galaxy. 

On the modeling side, uncertainties remain in the treatment of very young populations (the nebular emission) and in older populations (e.g. TP-AGB stars). The use of an MCMC fitting method reveals the blunt truth about these uncertainties. We are unable to fit parameters such as metallicity, SFR, and old ages. On the other hand, the three simple parameters of young population age, A$_V$, and stellar mass are relatively well constrained. This is due to the fact that the young age is mainly determined from the Balmer break, the A$_V$ mainly from the UV slope, and the stellar mass from the overall normalisation of the SED, three parameters which are well constrained in the SEDs fitted here. Overall, the constraints on these parameters should be robust.

The ages of the galaxies presented here are very young, and based on the appearance of many of them (including several grand spiral galaxies) it is clear that several, if not all, must have a longer history of star formation. Unfortunately, any older stellar populations are hidden due to the brightness of the younger stellar population, a bias that is true for observers of LBGs at both at low and high redshift. Hence, the ages presented here are merely a measure of the latest burst of star formation. Seen from the perspective of studies of LBG galaxies, the ages agree very well with previous results, especially those of very high redshift surveys from e.g.~Verma et al.~(2007) or Yabe et al.~(2009). As LBGs are selected based on their bright UV colours they are expected to be strongly star forming galaxies in the high redshift Universe. The young population ages consistently found in SED fitting confirms this expectation. The dust contents are slightly higher than the values found at high redshift; here we find dust extinctions of A$_V \approx 0.5 - 2.0$~mag, compared to a median dust extinction of A$_V \approx 0.7 - 1.2$~mag in the $z\sim2.5$ results of Shapley et al.~(2001) and Papovich et al.~(2001) or of A$_V \approx 0.3$~mag in the $z\sim5$ results of Verma et al.~(2007). This also seems reasonable considering the potentially longer evolution time. Masses are similar at both high and low redshift, with the lower redshift galaxies having a larger spread going up to masses of $10^{11}$~M$_\odot$. 

Overall, the properties of $z \sim 1$ LBGs are very similar to those of higher redshift LBGs. This is contrary to what has recently been seen in another large group of high redshift galaxies, the Ly$\alpha$ emitters (LAEs). For LAEs, it has been reported that lower redshift LAEs have different properties to higher redshift LAEs (Finkelstein et al.~2009, Nilsson et al.~2009a, Guaita et al.~2010, Nilsson et al.~2010). In these publications, it has been shown that galaxies selected through the Ly$\alpha$ technique at lower redshifts tend to be more massive and dustier than those selected at higher redshifts. The reason for this evolution is expected to be due to the fact that selecting Ly$\alpha$ emitting galaxies will select different objects at different times, due to the number density evolution of objects capable of Ly$\alpha$ emission across time. However, whereas Ly$\alpha$ can be produced by many sources, the strength of the Lyman break of a galaxy will only depend on the star formation history of the galaxy and the absorption of the intergalactic medium. This could be the reason why LBGs exhibit similar properties across time, as opposed to LAEs. Hence, the selection methods finding LBGs and LAEs are in some ways orthogonal, in that the LBG technique will find the same galaxies at all redshifts, and the LAE method will find a representation of all Ly$\alpha$ emitting objects in each redshift slice it surveys.

\subsection{The importance of the grism photometry}
\begin{figure}[!ht]
\begin{center}
\epsfig{file=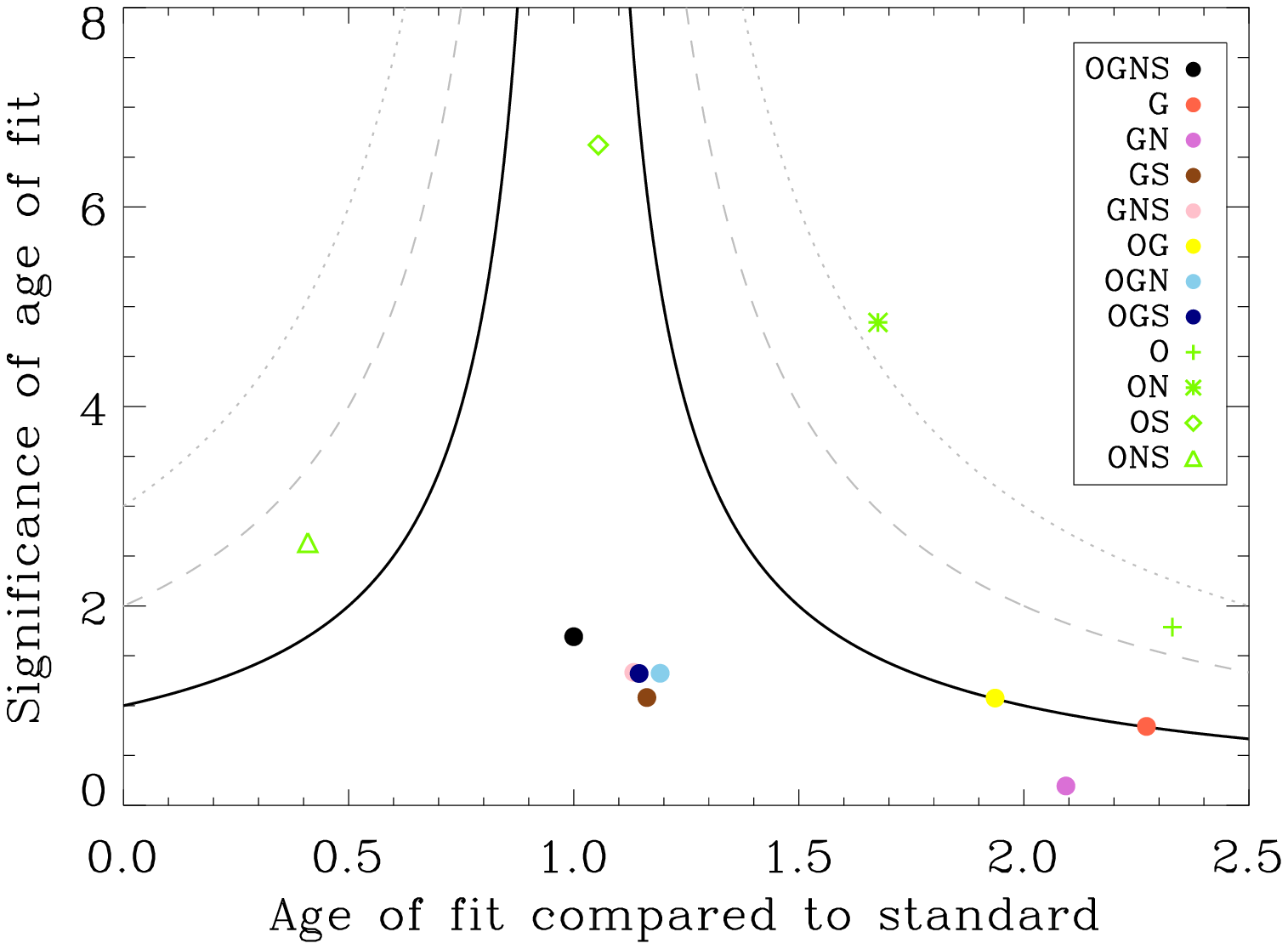,width=9cm}
\epsfig{file=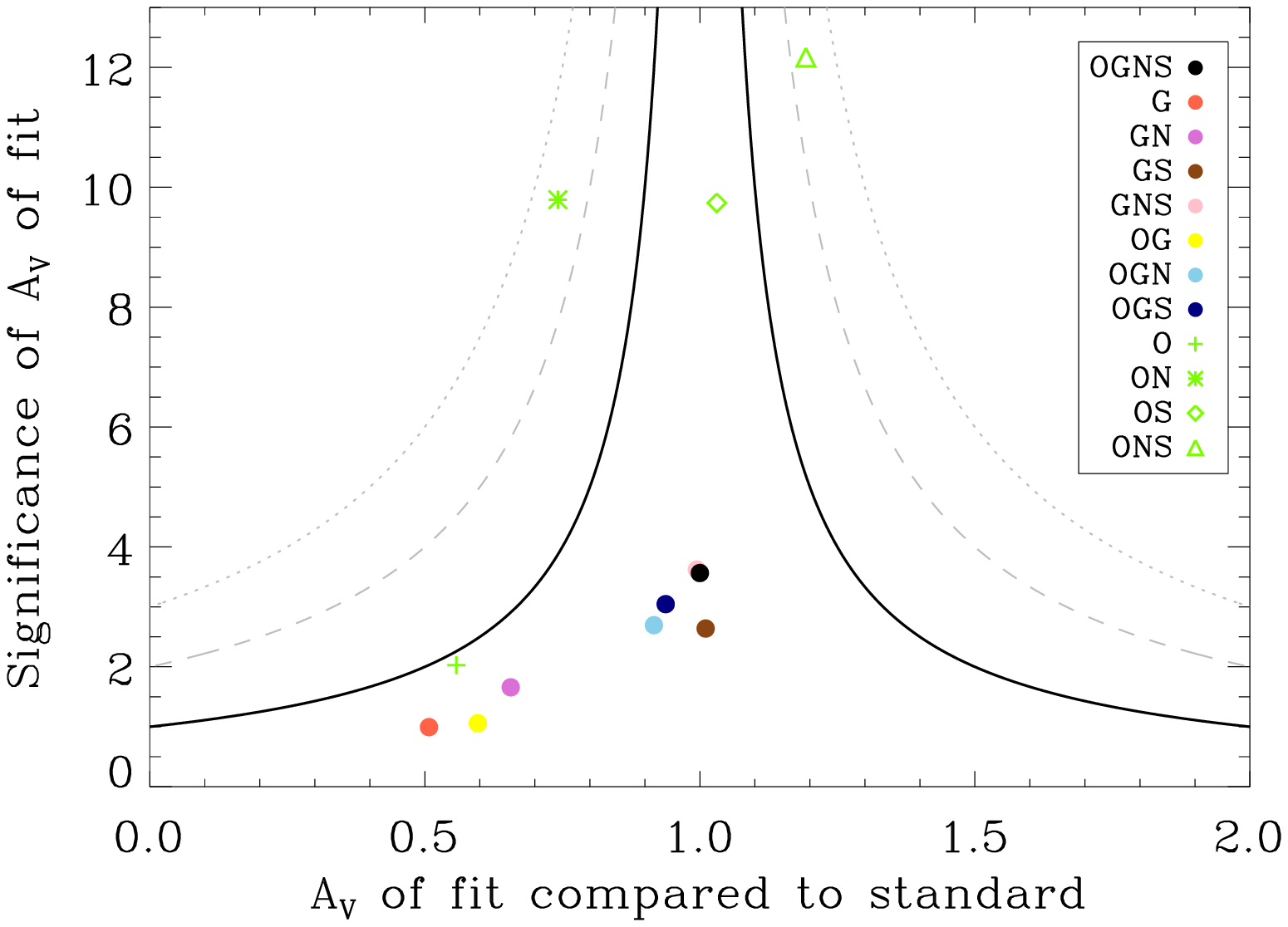,width=9cm}
\epsfig{file=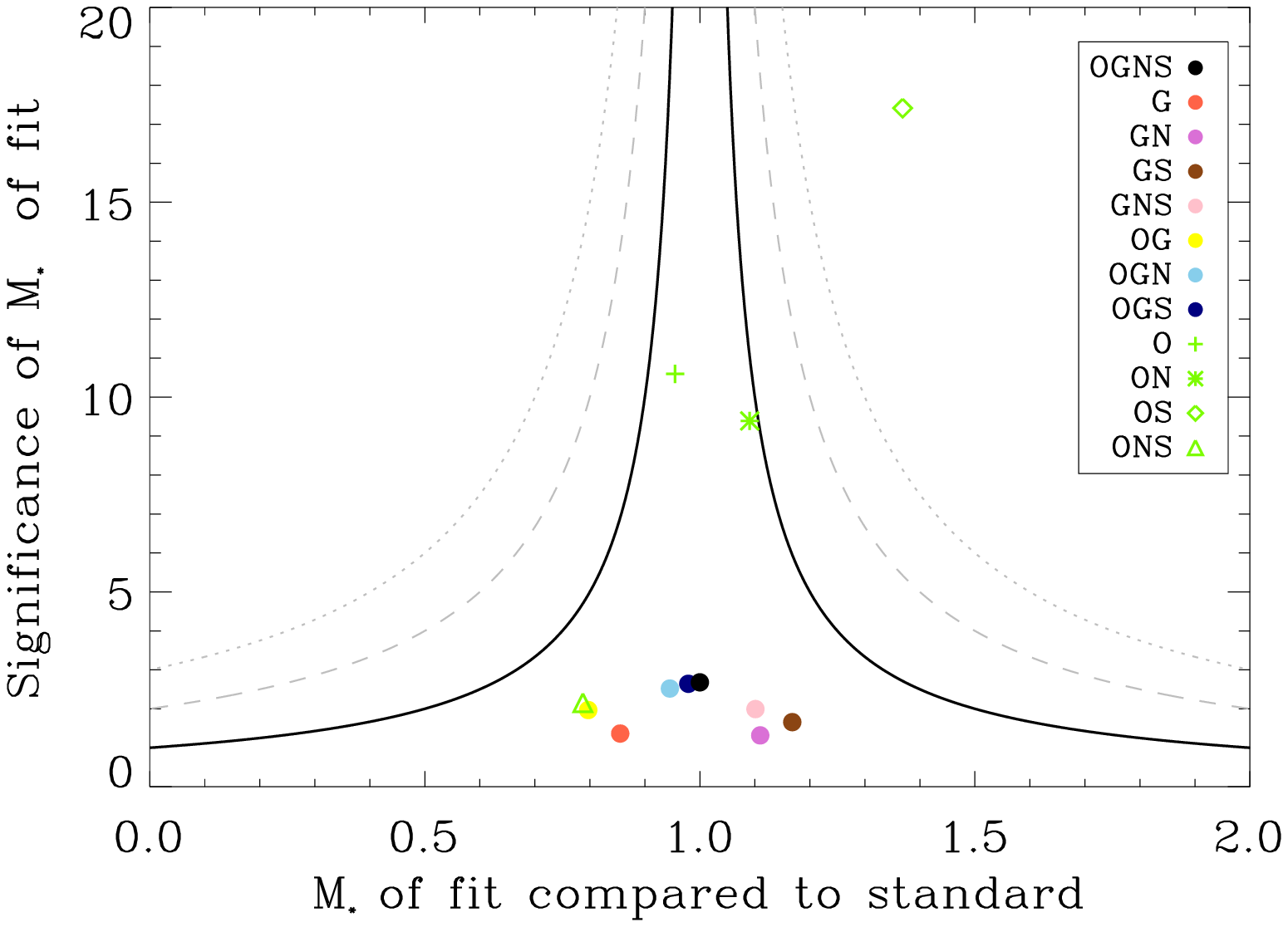,width=9cm}
\caption{Test of the importance of various photometric bands for the fitting. The three plots are shown for the parameters age, A$_V$ and stellar mass, from top to bottom. The \emph{x}-axes show the value for each fit normalised to the value obtained with all filter passbands used (OGNS, black points). The \emph{y}-axes show the value in each fit divided by the error bar on that value, indicating how well determined the value is. The various passbands included are UV/optical (``O''), grism (``G''), near-infrared (``N'') and Spitzer/Ch1 (``S''). The solid, dashed, and dotted line indicates the area within which points are within $1$,~$2$,~and~$3\sigma$ of the OGNS value.  }
\label{fig:phottest}
\end{center}
\end{figure}

To test the importance of the grism spectral points in the fits, we also ran the SED fits for various combinations of photometric points included. For this test, LBGs\_\# 59257, 63984, 66235 and 69919 were selected as test cases as they represent a range in $\chi^2$, morphology and number of grism points. The SEDs were divided into four parts; the UV/optical part (``O''), the grism points (``G''), the near-infrared points (``N'') and the single Spitzer/IRAC Ch1 point (``S''). New SED runs were then made with realistic permutations of these photometric parts. The results for one object, LBG\_69919,  are presented in Fig.~\ref{fig:phottest}.   
For the analysis of these results, the values of the fit including all bands (OGNS) are assumed to be the most correct, standard value. In Fig.~\ref{fig:phottest}, both the accuracy of the fit, in comparison to the OGNS result, and the significance of the result are shown. Points within the solid line cone all have parameter results that agree with the OGNS result within $1\sigma$. Fits \emph{without} the grism photometry included are all shown with green symbols. 

Several conclusions can be drawn from this Figure and the results of the fits of the other three objects. Firstly, most of all fits including the grism points are within $1\sigma$ of the OGNS values, whereas the fits without the grism points often lie far from the standard values. Thus, the grism points are necessary to get the best possible result. Another conclusion, and also part of the reason why the fits with grism points are always within $1\sigma$, is that the error bars are typically larger on the fits with grism points, i.e. a lower significance of the parameter. This can be slightly surprising, considering that the $\chi^2_r$ are consistent throughout all fits (e.g. it varies between $0.8 - 3.1$ for LBG\_69919). One possible reason for this is that the stellar templates used are not accurate enough in predicting the Balmer break in the galaxies. With up to 68 photometric points determining the Balmer break, albeit with large error bars, small uncertainties in the spectral population modeling could potentially cause large uncertainties in the fitted parameters. 

Further, fits including the Spitzer/IRAC Ch1 point are typically better fit than those without. An important conclusion is thus that it is always important to sample also the restframe near-infrared when fitting stellar populations. Finally, in Fig.~\ref{fig:phottest} it is seen that fitting the SEDs consisting of only the grism points reveals parameter results that are within $1\sigma$ envelope of the OGNS results. The significance of the results in this case are not very high, but the parameters are constrained enough to give interesting results, also in studies where other multi-wavelength data is not available. This conclusions seems to break down though, as fewer grism data-points are available, as in the case of e.g. LBG\_59257 or LBG\_66235.

Note though, that the conclusions presented here are valid for galaxies at $z \sim 1$ where the grism points sample the Balmer and 4000~{\AA} breaks of the galaxies. It is unlikely that fitting only grism points would give good results for other redshifts, although it cannot be excluded from the present data.

\subsection{Conclusion}
To conclude, we have demonstrated that including grism spectrophotometric points in SED fitting of $z\sim1$ LBGs have improved the fits of these galaxies greatly. The $z\sim1$ LBGs all have very young ages, possibly with one or more underlying older populations outshone from the brighter younger population. In order to be able to constrain older populations better, more accurate photometry in the restframe near-infrared bands (observed Spitzer bands) would be necessary, which is challenging for galaxies at $z \sim 1$ and even more difficult for higher redshift galaxies with existing observatories. The LBGs have moderate to high dust extinction and the masses are similar to those of higher redshift LBGs. Considering the similarity of properties of LBGs at higher and lower redshifts, it appears that selecting high redshift galaxies according to their UV properties selects the same type of galaxy, irrespective of redshift, as opposed to the selection via e.g.~Ly$\alpha$ emission. 

\begin{thebibliography}{99}
\bibitem{balestra}Balestra, I., Mainieri, V., Popesso, P., et al., 2010, A\&A, 512, A12
\bibitem{Balogh}Balogh, M.L., Morris, S.L., Yee, H.K.C., Carlberg, R.G., \& Ellingson, E., 1999, ApJ, 527, 54
\bibitem{Bolzonella}Bolzonella, M., Miralles, J.-M., Pell{\'o}, R., 2000, A\&A 363, 476
\bibitem{Bruzual} Bruzual G.A., Charlot S., 2003, MNRAS, 344, 1000
\bibitem{bunk}Bunker, A.J., Stanway, E.R., Ellis, R.S., \& McMahon, R.G., 2004, MNRAS, 355, 374
\bibitem{burga}Burgarella, D., Le Floc'h, E., Takeuchi, T.T., et al., 2007, MNRAS, 380, 986
\bibitem{calzetti}Calzetti, D., Armus, L., Bohlin, R.C., et al., 2000, ApJ, 533, 682
\bibitem{Daddi}Daddi, E., Cimatti, A., Renzini, A., et al., 2004, ApJ, 617, 746
\bibitem{finkel}Finkelstein, S.L., Cohen, S.H., Malhotra, S., \& Rhoads, J.E., 2009, ApJ, 700, 276  
\bibitem{franx}Franx, M., Labb\'e, I., Rudnick, G., et al., 2003, ApJ, 587, L79
\bibitem{Fynbo02}Fynbo, J.P.U., M{\o}ller, P., Thomsen, B., et al., 2002, A\&A, 388, 425 
\bibitem{gaw06}Gawiser, E., Van Dokkum, P.G., Herrera, D., et al., 2006, ApJS, 162, 1
\bibitem{goods}Giavalisco, M., Ferguson, H.C., Koekemoer, A.M., et al., 2004, ApJ, 600, L93
\bibitem{guaita}Guaita, L., Gawiser, E., Padilla, N., et al., 2010, ApJ, 714, 255
\bibitem{gunn}Gunn, J.E., \& Stryker, L., ApJS, 52, 121
\bibitem{Hamil}Hamilton, D., 1985, ApJ, 297, 371
\bibitem{Kauffmann}Kauffmann, G., Heckman, T.M., White, S.D.M., et al., 2003, MNRAS, 341, 33
\bibitem{kenni2}Kennicutt, R.C., 1998, ARAA, 36, 189
\bibitem{Kriek}Kriek, M., van Dokkum, P.G, Franx, M., et al., 2006, ApJ, 645, 44 
\bibitem{lefevre}Le F{\`e}vre, O., Vettolani, G., Garilli, B., et al., 2005, A\&A, 439, 845
\bibitem{Leit}Leitherer, C., Schaerer, D., Goldader, J.D., et al., 1999, ApJS, 123, 3
\bibitem{Luo}Luo, B., Bauer, F.E., Brandt, W.N., et al., 2008, ApJS, 179, 19
\bibitem{mignoli}Mignoli, M., Cimatti, A., Zamorani, G., et al., 2005, A\&A, 437, 883
\bibitem{MW1993}M{\o}ller, P., \& Warren, S.J. 1993, A\&A 270, 43
\bibitem{nilsson07}Nilsson, K.K., M{\o}ller, P., M{\"o}ller, O., et al., 2007, A\&A, 471, 71
\bibitem{nilsson09}Nilsson, K.K., Tapken, C., M{\o}ller, P., et al., 2009a, A\&A, 498, 13
\bibitem{nilsson09b}Nilsson, K.K., M{\"o}ller-Nilsson, O., M{\o}ller, P., Fynbo, J.P.U., \& Shapley, A.E., 2009b, MNRAS, 400, 235
\bibitem{nilsson10}Nilsson, K.K., {\"O}stlin, G., M{\o}ller, P., et al., 2010, submitted to A\&A, arXiv:1009.0007
\bibitem{nonino}Nonino, M., Dickinson, M., Rosati, P., et al., 2009, ApJS, 183, 244
\bibitem{ouchi04}Ouchi, M., Shimasaku, K., Okamura, S. et al., 2004, ApJ, 611, 660
\bibitem{overz}Overzier, R.A., Heckman, T.M., Tremonti, C., et al., 2009, ApJ, 706, 203 
\bibitem{papo}Papovich, C., Dickinson, M., \& Ferguson, H.C., 2001, ApJ, 559, 620
\bibitem{petti}Pettini, M., Shapley, A.E., Steidel, C.C., et al., 2001, ApJ, 554 981
\bibitem{pirz}Pirzkal, N., Xu, C., Malhotra, S., et al., 2004, ApJ, 154, 501
\bibitem{Ranalli}Ranalli, P., Comastri, A., \& Setti, G., 2003, A\&A, 399, 39
\bibitem{ravi}Ravikumar, C.D., Puech, M., Flores, H., et al., 2007, A\&A, 465, 1099
\bibitem{Retzlaff}Retzlaff, J., Rosati, P., Dickinson, M., et al., 2010, A\&A, 511, A50 
\bibitem{shapley01}Shapley, A.E., Steidel, C.C., Adelberger, A.E., et al., 2001, ApJ, 562, 95
\bibitem{shapley}Shapley, A.E., Steidel, C.C., Pettini, M., \& Adelberger, K.L., 2003, ApJ, 588, 65
\bibitem{steidel96}Steidel, C.C., Giavalisco, M., Pettini, M., Dickinson, M., \& Adelberger, K.L., 1996, ApJ, 462, L17
\bibitem{Steidel}Steidel, C.C., Adelberger, K.L., Giavalisco, M., Dickinson, M., \& Pettini, M., 1999, ApJ, 519, 1
\bibitem{straughn}Straughn, A.N., Pirzkal, N., Meurer, G. R., et al., 2009, AJ, 138, 1022
\bibitem{vanz}Vanzella, E., Cristiani, S., Dickinson, M., et al., 2008, A\&A, 478, 83
\bibitem{vene07}Venemans, B.P., R{\"o}ttgering, H.J.A,; Miley, G.K., et al., 2007, A\&A, 461, 823
\bibitem{verma}Verma, A., Lehnert, M.D., F{\"o}rster-Schreiber, N.M., Bremer, M.N., \& Douglas, L., 2007, MNRAS, 377, 1024
\bibitem{xu}Xu, C., Pirzkal, N., Malhotra, S., et al., 2007, AJ, 134, 169
\bibitem{yabe}Yabe, K., Ohta, K., Iwata, I., et al., 2009, ApJ, 693, 507
\end{thebibliography}
\end{document}